\documentclass[]{spie}  
\pdfoutput=1
 
\usepackage{amsmath,amsfonts,amssymb}
\usepackage{graphicx}
\usepackage{multirow}
\usepackage[colorlinks=true, allcolors=blue]{hyperref}
\usepackage{rotating}

\title{End-to-end Simulation of the SCALES Integral Field Spectrograph}

\author[a]{Zackery Briesemeister}
\affil[a]{University of California, Santa Cruz, Santa Cruz CA 95060}
\author[b]{Steph Sallum}
\affil[b]{University of California, Irvine, Irvine CA 92697}
\author[a]{Andrew Skemer}
\author[a]{R. Deno Stelter}
\author[a]{Philip Hinz}
\author[c]{Timothy Brandt}
\affil[c]{University of California, Santa Barbara, Santa Barbara CA 93106}

\authorinfo{Further author information: (Send correspondence to Zackery Briesemeister)\\E-mail: zbriesem@ucsc.org}

\begin{document} 
\maketitle

\begin{abstract}

We present end-to-end simulations of SCALES, the third generation thermal-infrared diffraction limited imager and low/med-resolution integral field spectrograph (IFS) being designed for Keck. The 2-5 micron sensitivity of SCALES enables detection and characterization of a wide variety of exoplanets, including exoplanets detected through long-baseline astrometry, radial-velocity planets on wide orbits, accreting protoplanets in nearby star-forming regions, and reflected-light planets around the nearest stars. The simulation goal is to generate high-fidelity mock data to assess the scientific capabilities of the SCALES instrument at current and future design stages. The simulation processes arbitrary-resolution input intensity fields with a proposed observation pattern into an entire mock dataset of raw detector read-out lenslet-based IFS frames with calibrations and metadata, which are then reduced by the IFS data reduction pipeline to be analyzed by the user.

\end{abstract}


\section{INTRODUCTION}
\label{sec:intro} 

The current generation of high-contrast coronagraph-assisted imagers and (integral field) spectrometers have succeeded in identifying and characterizing a population of massive, wide-period exoplanets by bulk, chemical, and atmospheric properties in exceptional detail. While this limited population has been an excellent probe of this specific region of parameter space, an extended exploration would access the outcomes of various planet formation and migration scenarios. The goal of accessing a larger region of parameter space of possible exoplanets motivates the development of new instruments that improve on previous designs. This includes use of advanced adaptive optics combined with coronagraphic integral field spectroscopy mounted on large/extremely-large telescopes that extends to longer wavelengths (2-5 $\mu$m), where self-luminous exoplanets have greatest contrast with their host star. These longer wavelengths have been otherwise inaccessible given the high background radiation, detector sensitivity and instrument design.

The precursor instrument Arizona Lenslets for Exoplanet Spectroscopy \cite{2015SPIE.9605E..1DS} (ALES) in the Large Binocular Telescope Interferometer (LBTI) \cite{2008SPIE.7013E..28H, 2012SPIE.8445E..0UH, 2014SPIE.9146E..0TH} has provided the first spatially-resolved thermal-infrared spectral characterization of substellar companions \cite{2019AJ....157..244B, 2020AJ....160..262S}, and Santa Cruz Array of Lenslets for Exoplanet Spectroscopy (SCALES) will improve on ALES in optomechanical design and sensitivity. While ALES bridges this technological \cite{2018SPIE10702E..3LH, 2018SPIE10702E..3FS, 2018SPIE10702E..0CS} and scientific gap, the dedicated instrument SCALES vastly improves upon stability and sensitivity of the nascent technology in order to provide 10-m class diffraction-limited thermal infrared low-/med-resolution spectra and imaging. This paper describes the end-to-end simulation of SCALES,  accompanied by the implementation of the data reduction and analysis tools repurposed and improved upon from the ALES data reduction pipeline \cite{2018SPIE10702E..2QB}.

\section{SCALES at a Glance}

\begin{figure}[h]
\includegraphics[angle=-90, origin=c, width=16cm, trim={7cm 0 27cm 0}]{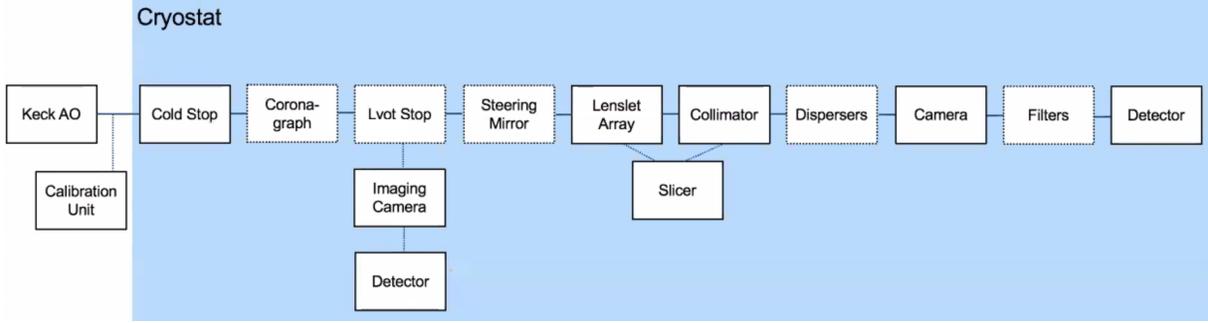}
\centering
\caption{Schematic of the SCALES optical layout. Keck AO delivers f/15 light through the entrance window. For the imager, light is relayed to a dedicated detector. For the spectrograph, the light passes through the focal plane coronagraph and Lyot stop, forming an image at the lenslet array. The low-resolution mode comprises the lenslet array, collimator, disperser and camera. The med-resolution mode selects a $18\times18$ spaxel region from the lenslet array and relays the light through an image slicer that reorients the spectra in a pseudoslit that is then dispersed and imaged on the detector. The calibration unit is described here \cite{2018SPIE10702E..2QB}.}
\label{fig:fig1}
\end{figure}

SCALES comprises a low-resolution integral field spectrograph and a medium resolution integral field spectrograph that share coronagraphic foreoptics transmissive from 2-5 $\mu$m, and a 1-5 $\mu$m imager. A complete description of the current optomechanics of SCALES is presented in these proceedings by Stelter et al. in paper \#11447-110. Figure \ref{fig:fig1} depicts the schematic of the SCALES optical layout for the three modes.

For the low-resolution mode, f/15 light from Keck AO enters the cryogenic dewar through a CaFl entrance window and is relayed through a fixed cold-stop followed by a linear slide containing focal plane vector-vortex coronagraphs optimized for $K$, $L$, and $M$ bands. The light is magnified by 22.8 by an off-axis-ellipse with an intermediate pupil plane that contains Lyot stops, pupil apodizers and non-redundant pupil masks. The light is relayed to the 108$\times$108 square lenslet array, where the 341 micron, f/8 square lenslets sample the  $2.15^{\prime\prime}\times2.15^{\prime\prime}$ field with a lenslet pitch of $0.02^{\prime\prime}\times0.02^{\prime\prime}$ to Nyquist sample at 2.0 $\mu$m. A two-element collimator produces a pupil plane at the disperser wheel, which contains reflective double-pass LiF prism with gold coating on the second surface for each bandpass. A two-element camera images the spectra onto the Teledyne HAWAII-2RG \cite{2008SPIE.7021E..0HB} detector.

For the med-resolution mode, a piezo-controlled tip-tilt stage directs the astrophysical scene to a subarray of $18\times18$ spaxels on the side of the lenslet array at the same magnification ($0.36^{\prime\prime}\times0.36^{\prime\prime}$ field of view with a lenslet pitch of $0.02^{\prime\prime}\times0.02^{\prime\prime}$). The slicer system rearranges the lenslet pupil images back into the spectrograph as a single pseudo-slit where a pupil forms at gratings designed to disperse the spectra without order overlap, and imaged onto the detector similarly.

The imager will pick light off from the Lyot stop wheel towards a dedicated camera and detector system with a field of view of $20^{\prime\prime}\times20^{\prime\prime}$ at a plate scale of $0.01^{\prime\prime}\times0.01^{\prime\prime}$. The imager has NIRC2-like filters.

\begin{table}

\begin{center}
\begin{tabular}{ |p{2cm}||p{2cm}|p{2cm}|p{2cm}|p{2cm}  }
 \hline
 \multicolumn{4}{|c|}{SCALES Spectral Summary} \\
 \hline
  & Wavelengths  & Band & Resolution\\
 \hline
    \multirow{6}{6em}{Low-Resolution Spectroscopy} & 2.0-3.7 $\mu$m & \text{water ice} & $\sim$ 60  \\ 
    & 2.0-2.4 $\mu$m & \text{$K$ Band} & $\sim$ 200 \\ 
    & 2.0-5.0 $\mu$m & \text{SEDs} & $\sim$ 35 \\ 
    & 2.9-4.15 $\mu$m & \text{$L$ band} & $\sim$ 80  \\
    & 3.1-3.5 $\mu$m & \text{CH${}_{4}$ + PAH} & $\sim$ 250  \\
    & 4.5-5.2 $\mu$m & \text{$M$ band} & $\sim$ 140  \\
     \hline
    \multirow{3}{3em}{Medium-Resolution Spectroscopy} & 2.0-2.4 $\mu$m & \text{$K$ band} & $\sim$ 5000  \\ 
    & 2.9-4.15 $\mu$m & \text{$L$ band} & $\sim$ 3500  \\ 
    & 4.5-5.2 $\mu$m & \text{$M$ band} & $\sim$ 7000  \\
 \hline
\end{tabular}
\end{center}
\caption{Wavelength bands with filters in SCALES for the low-resolution and medium-resolution integral field spectrographs. }
\end{table}

\section{Simulation Architecture}

SCALES is currently in the preliminary design stage. The simulation tool is tailored towards investigating the impact of optomechanical design choices on the fiducial science cases for SCALES, enumerated in paper \#11447-110. Simulation of the outputs from instruments is necessary to predict their performance. The fidelity of such simulations is limited by the reproduction of the physics and astrophysics of the observation process, optomechanical design, and reduction prescription that can be interpreted in finite computation time. 

\begin{sidewaysfigure}
\includegraphics[width=24cm, trim={2cm 3cm 0cm 5cm}]{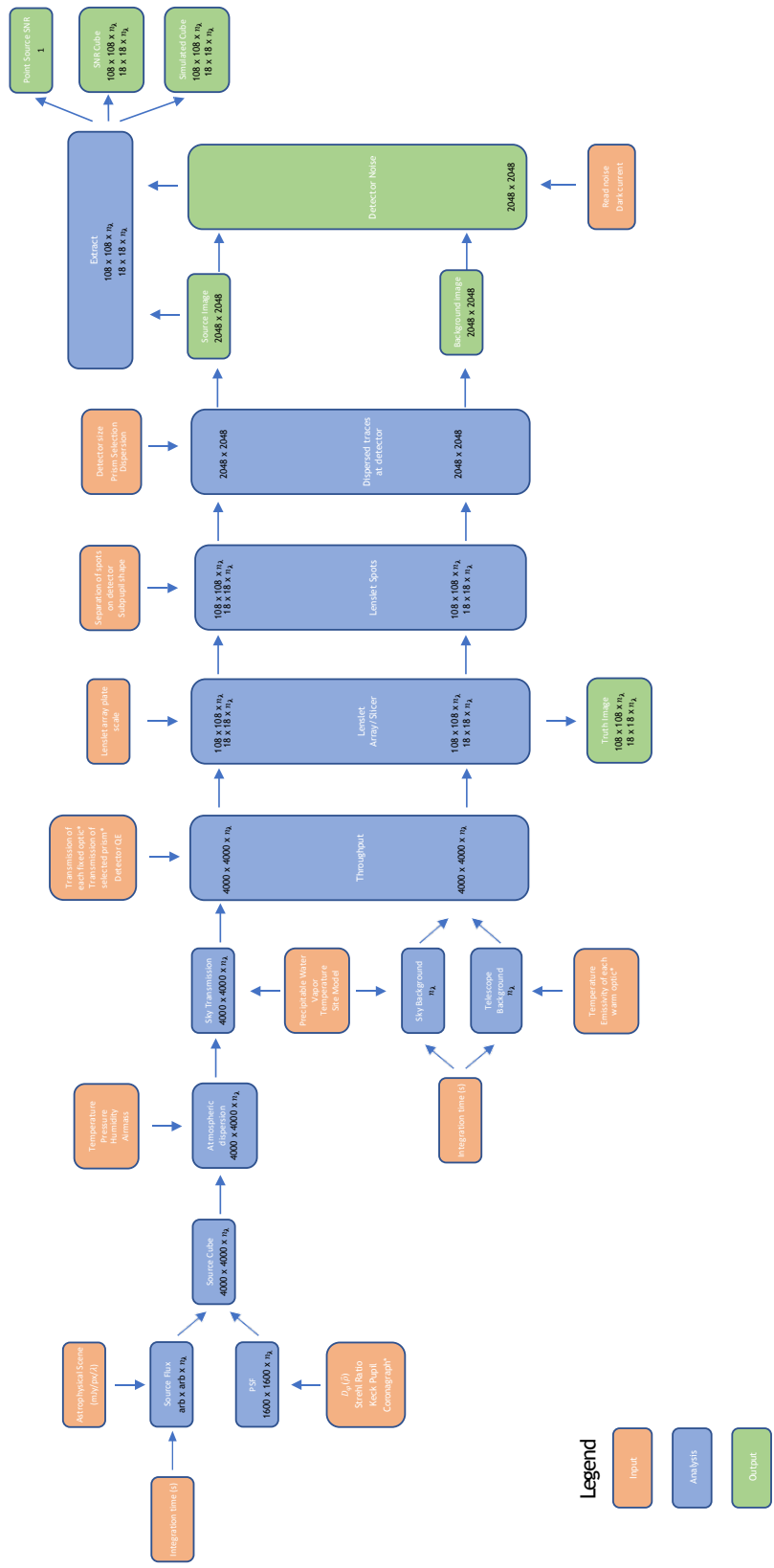}
\centering
\caption{A schematic of the forward-direction of the simulation necessary to reproduce the astrophysics, tellurics and hardware of SCALES observations. For a given integration time, PSF instances are convolved with astrophysical scenes and deflected by the atmospheric dispersion. Their intensity and sampling are modulated by the transmission and lenslet array, respectively. The low-/med-resolution modes reformat the image according to their spectrograph focal plane, and result in detector images that are reduced by the data reduction pipeline. The black numbers describe internal dimensions used in the simulation for the top/low-resolution and bottom/med-resolution modes, respectively. The coronagraph and assessment of transmissions are marked with an asterisk *, denoting incomplete implementation because they are currently being designed.} 
\label{fig:fig2}
\end{sidewaysfigure}

For these SCALES simulations, astrophysical fields are provided by the user in the form of a spatiospectral data cube in mJy/px/$\lambda$ at higher spatial and spectral resolution than the instrument, along with metadata, including sampling rate, position, date/time of observation, duration, and atmospheric conditions. The simulation will generate a mock dataset from an entire low-/med-resolution IFU observation in detector readout units and/or file facsimiles. The simulations for the imager are not currently included. This tool was written in Python. In this section, we describe the models used to reproduce the physical processes of imaging spectroscopy in SCALES. Figure \ref{fig:fig2} depicts how the optomechanics from Figure \ref{fig:fig1} are translated into software.

\subsection{Adaptive Optics Simulation}

While the point-spread function of the instrument is an important component to reproduce with fidelity, the nature of the point-spread function at the focal planes of the imager and lenslet-array remains a point of active development. The large wavelength sensitivity merits having three focal plane coronagraphs optimized to $K$, $L$, and $M$ bands, respectively, which are also being developed and are not implemented in the simulation yet.

Any astrophysical intensity field observed from the ground is modulated by the Earth's atmosphere and can be partially spatiotemporally corrected by an adaptive optics system. Simulating one long exposure sampling thousands of correlated phase screens is computationally prohibitive. Our approach is to use a simple analytical model to express images of finite exposure times, eschewing the time-correlation of phase screens in favor of faster computation. Given a user-chosen (default von-K\'arm\'an, $L_0 = 8 $ m) structure function of residual phase, $D_{\phi}(\vec{\rho})$, a point spread function averaged over exposure-time $T$ for turbulence lifetime $\tau$ will have intensity $\langle I \rangle_T = \mathbb{E}(I) + \sigma$, such that $var(\sigma) = \frac{\tau}{T}(\mathbb{E}(I^2) - \mathbb{E}(I)^2)$. The intensity term $I$ is expressed in terms of the associated optical transfer function (OTF), $\hat{h}(\vec{\rho}/\lambda) = \hat{h}_A(\vec{\rho}/\lambda)\hat{h}_T(\vec{\rho}/\lambda)$, for atmosphere OTF $\hat{h}_T(\vec{\rho}/\lambda) = exp({-\frac{1}{2}D_{\phi}(\vec{\rho})})$, telescope OTF $\hat{h}_T(\vec{\rho}/\lambda) \propto \iint P(\vec{r})P(\vec{r} + \vec{\rho})d\vec{r}$, and $P$ entrance pupil transmission function. Individual point spread functions are drawn from this model for each frame.

The final PSF is obtained by multiplying the OTF due to high order effects with ideal correction on low order modes, from low order wavefront errors, and due to uncertainties from the optical system and instrument set by design requirements of SCALES. The final OTF is tuned such that the Strehl ratio at 3.7 $\mu$m is 0.85 to reproduce Keck AO-like optical quality.

Given the extended wavelength coverage of SCALES, departure from zenith will result in atmospheric deflection of the position of a point-source as a function of wavelength. This uncorrected deflection is simulated by affine transformation along the direction perpendicular to the horizon with magnitude associated with a given temperature, pressure, relative humidity and airmass \cite{2007JOptA...9..470M}. While the effect of relative astrometry is suppressed by the integral field spectrograph, the offset has important implications on the hardware and observation modes we employ.

\subsection{Transmission and Background}

The column of atmosphere in the direction of the source region will modulate the transmission of astrophysical photons from the source, modeled by the theoretical sky background available from the Gemini observatory calculated with the atmospheric model ATRAN \cite{1992nstc.rept.....L}.

We use the theoretical sky background available from the Gemini Observatory, which includes the sky transmission calculated from with the atmospheric model ATRAN \cite{1992nstc.rept.....L}, a 273 K continuum to simulate the sky. The telescope background and AO system background are modeled as blackbodies at 273 K and their emissivity in the thermal infrared is currently roughly estimated at $\sim35\%$.

Each gold-coated mirror and transmissive optic is modelled using emissivities and transmission profiles from the manufacturer, respectively. The lenslet array is photolithography-etched silicon with imperfections on the sub-micrometer scale, which is ignored. The transmission of the lenslet array is estimated at 98\%. 

\subsection{Simulating the Spectrographs}

\begin{figure}[h]
\includegraphics[angle=-90, origin=c, width=16cm, trim={5cm 0 20cm 0}]{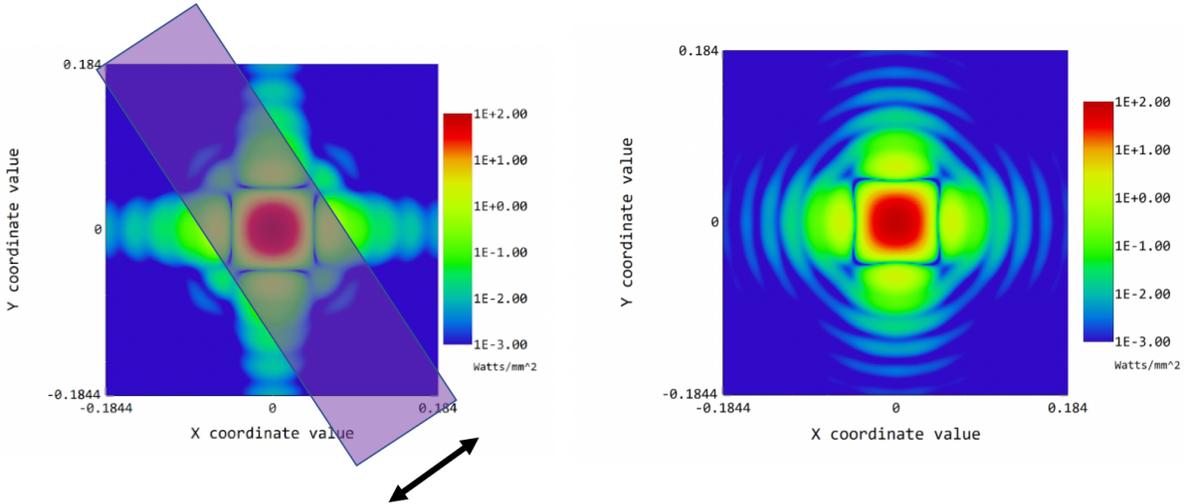}
\centering
\caption{Left: An image of the 5.0 micron lenslet pupil caused by diffraction comprising the flux incident on f/8 lenslet of the lenslet array. The purple box denotes the direction of dispersion for low-resolution mode. Right: An image of the 5.0 micron lenslet pupil caused by diffraction of the square f/8 lenslet and the pinhole placed at the geometric focus of the lens. The pinhole throughput is $\sim83\%$ at 5 microns.}
\label{fig:fig3}
\end{figure}

At the focal plane of each lenslet, an image of the exit lenslet pupil will form, comprising of all the light from the image incident to the spatial extent of the lenslet. In the optical design we place a pinhole grid of circles at a distance optimized by Zemax Physical Optics from the lens to suppress the square diffraction effects of each lenslet (Figure \ref{fig:fig3}). We determined with Zemax that the pinhole grid is preferred over not having one due to the dispersion direction and spectral resolution. The lenslet/pinhole PSF is calculated at each wavelength, and is scaled in amplitude by the input intensity.

In the case of the low-resolution spectrograph, the lenslet pupil images are then dispersed without spatial and spectral overlap, which forms images of dispersed lenslet pupils. For the med-resolution spectrograph, an $18\times18$ subset of lenslets are picked off and the lenslet pupil images are reordered by offset flat mirrors to the reflective disperser and relayed onto the detector.

\begin{figure}[h]
\includegraphics[width=16cm]{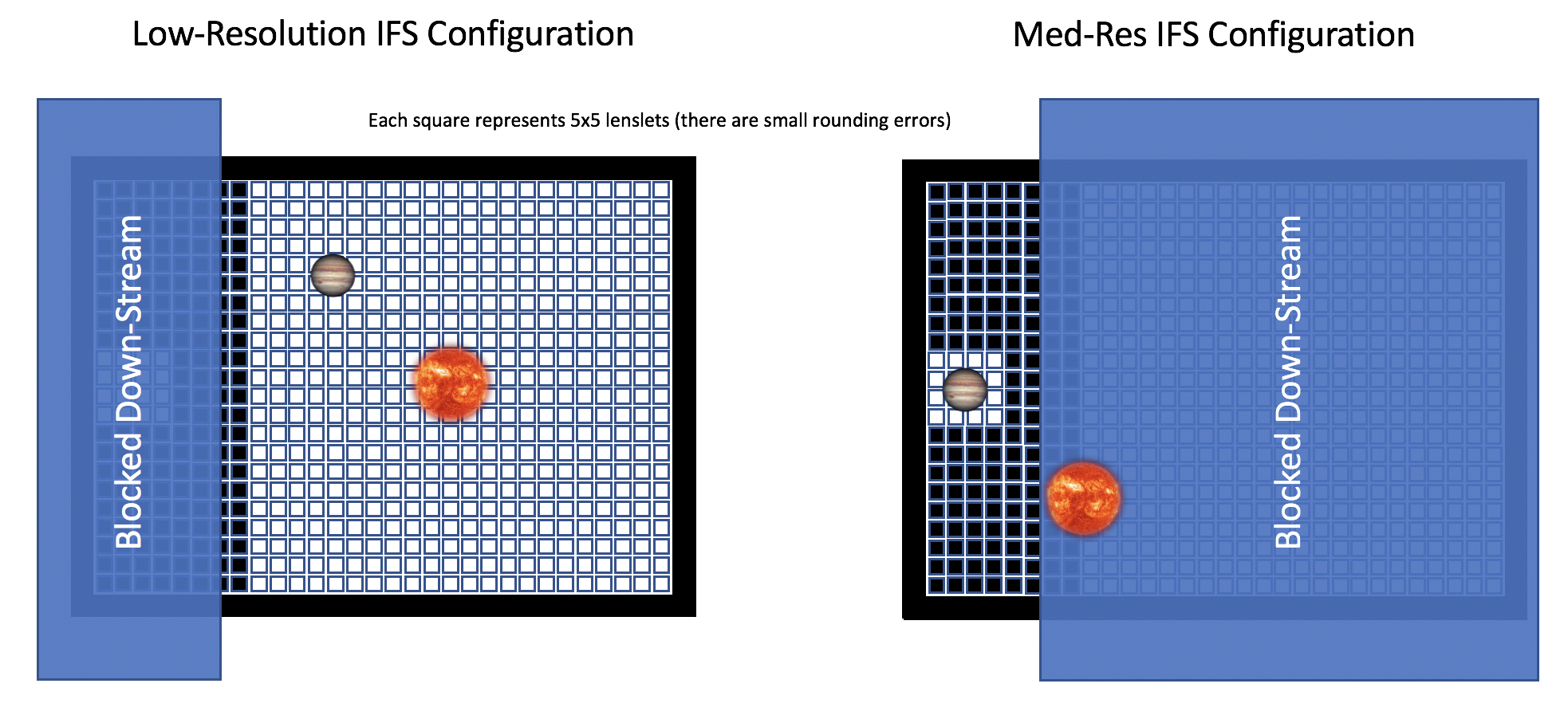}
\centering
\caption{Left: A cartoon for a star/exoplanet scene incident on SCALES at the low-resolution integral field spectrograph mode. The region of lenslets relayed to the disperser and detector are white in this image and decimated at a rate of 5 pixels. Right: A piezo-controlled mirror translates the star/exoplanet system in the med-resolution integral field spectograph mode, such that the exoplanet is incident on the active spaxels and the star is obstructed. The active spaxels are dispersed following the slicer pseudoslit to the detector in an array of $18\times18$ spectra containing spatiospectral information.} 
\label{fig:fig4}
\end{figure}

The mechanical distinction of the med-resolution mode compared to the low-resolution mode is depicted in Figure \ref{fig:fig4}. In low-resolution mode, the deviations of the grid of lenslet pupil images from rectilinear are characterized in the optomechanical design, and are applied as a 2D-polynomial shift that the data reduction pipeline is agnostic to. In med-resolution mode, predetermined shifts, set by the optomechanical design of the slicer, are applied to each spectrum.

Dispersion for low-resolution mode is performed by reflective double-pass LiF prism with gold-coated back and anti-reflection coated front tuned for each band pass. These prisms haven't been designed yet, so a linear dispersion profile is used to calculate the position at which monochromatic PSFs are imaged onto the detector. The final science image without noise comprises contributions of propagated monochromatic light at each wavelength, summed and binned at the detector pixel rate.

Dispersion for med-resolution mode is performed by gratings tuned for $K$, $L$, and $M$ bands. The position of monochromatic light is determined by the image slicer and a linear dispersion profile of the grating (as the gratings also have not been designed yet). The final science image without noise is then the contributions of propagated monochromatic light at each wavelength, summed and binned at the detector pixel rate.

\subsection{Detector Noise}
The detector for SCALES is a $2048\times2048$ pixel Teledyne HAWAII-2RG detector tuned to 5.3 micron cutoff. Generating high-fidelity realizations of detector noise requires complex modelling of correlated stationary and non-stationary noise components caused by solid-state physics. In order to approach fidelity, we modified the noise generator from (Rausher 2015) \cite{2015PASP..127.1144R} to apply to various readout methods and digitization processes. These modifications have been tested to reproduce the LBT/LMIRCam \cite{2010SPIE.7735E..3HS,2012SPIE.8446E..4FL} H2RG detector noise. 

For both modes, realizations of detector noise are added to the science frames. Calibration frames are also propagated through the same process. Apart from flux calibration, the calibration frames are necessary for the actual extraction step for the least-squares and sparse matrix extraction methods.

\subsection{Data Reduction Pipeline}

\begin{figure}[h]
\includegraphics[width=16cm, trim={0cm 0cm 0cm .35cm}]{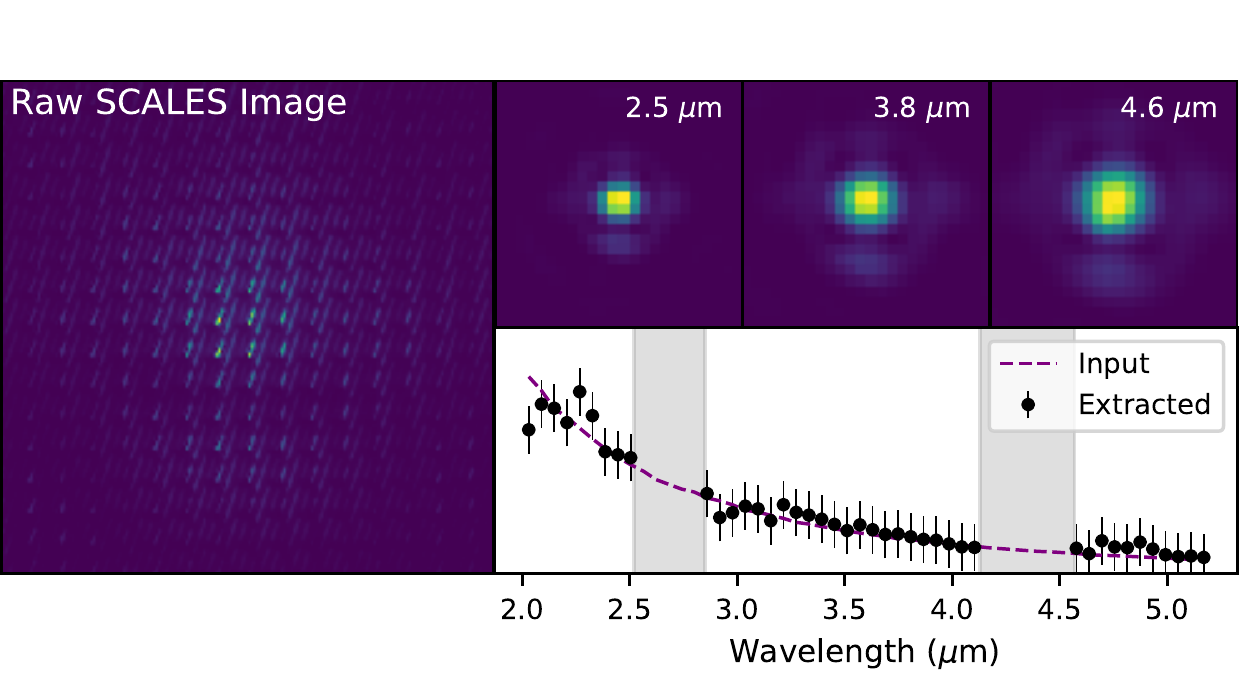}
\centering
\caption{Left: A zoomed-in single sky-subtracted data frame produce in low-resolution simulation of an A0 star at 10pc with the 2-5$\mu$m filter mode. The color stretch is square root in order to depict more spaxels. Right/Top: The extracted point spread function for the image at 2.5$\mu$m, 3.8$\mu$m, and 4.6$\mu$m, respectively. The color stretch is linear. Right/Bottom: The input and extracted spectrum with errors from 4 seconds on target and 4 seconds on sky performed with aperture extraction. The grey shading denote regions of high telluric absorption. Deviations in $K$ band are driven by red-leaks, warranting use of an informed extraction (least-squares) over aperture extraction.} 
\label{fig:fig5}
\end{figure}

In order to convert the simulated data in low-resolution mode to data cubes, the simulation uses $\texttt{MEAD}$ \cite{2018SPIE10702E..2QB} to facilitate the translation of raw detector readout frames and metadata into data cubes containing spatial and spectral information. This simulation tool has been extensively used to test the implementations of aperture, optimal \cite{1986PASP...98..609H}, least-squares \cite{2017JATIS...3d8002B, 2014SPIE.9147E..4ZD}, sparse matrix extraction of the spectra into data cubes, and we are exploring the efficiency of other data structures to optimize this extraction. Figure \ref{fig:fig5} depicts raw simulated data, the recovered images and injected/recovered spectra for a short integration of an A0 star at 10pc in $L$ band.

The med-resolution mode required development of a new module for $\texttt{MEAD}$, as the slicer pseudoslit reformats the dispersed $18\times18$ lenslet pupil images in a fundamentally different way than the lenslet array alone. The selected order of 324 spectra are separated by $\sim 6$ pixels and have the dynamic range limited by the detector layout of 2048 pixels, slicer design, and disperser profiles. These spectra are extracted with any of the four extraction methods previously implemented for the low-resolution mode for these longer spectra. This early data reduction pipeline for the med-resolution modes will drive investigation towards a more robust pipeline in the future. 

\section{Simulation and Design}

The simultaneous development of instrument and instrument simulation has many engineering and design benefits. The simulation has been used to quantify requirements set by the fiducial science cases and to test the consequences of update to the design of the instrument on meeting these requirements.

Background radiation sets unique limitations on the science capabilities of SCALES. The issue manifests most notably in the 2-5 $\mu$m filter in low-resolution mode, where the thermal background at the red end of the filter is orders of magnitude brighter than the thermal background at the blue end. For the desired fill factor of the detector, the spectra are limited in their separation, and the bright red ends of spectra neighbor the blue ends of other spectra. These red leaks are an inherent source of noise that would limit the usefulness of broad filters without appropriate filter design. This requirement has been investigated with the simulation and used to set the requirements on the transmission of light at the red end of this filter.

An incomplete list of the other uses of the simulation include the consequences of deviation from polynomial dispersion, deviation from rectilinear grid of lenslet pupil images at the focal plane, stability/reproducibility requirements of the disperser elements and other wheels, material transmission, drift, optical quality, etc. on target sources that we have identified as fiducial science cases for SCALES. 

Designing appropriate observation patterns is also relevant to maximize astrophysical photons while mitigating the thermal background at these wavelengths. The small field of view of the med-resolution mode when observing high-contrast targets necessitates an informed observation pattern to obtain calibrations, sky frames, dark frames and on target frames in a sky-rotating-frame, all of which we are using this tool to model and investigate.

\section{Astrophysics with SCALES}

The unique capabilities of SCALES opens an otherwise inaccessible parameter space of spatiospectral heterogeneity in astrophysics explored at small angular separation. SCALES is particularly capable of accessing unexplored regions of parameter space in which many detectable exoplanets are expected to exist. This is facilitated by combining thermal infrared (2-5 $\mu$m) sensitivity in the region of greatest exoplanet-star contrast \cite{2014ApJ...792...17S} and integral field spectroscopy for distinguishing exoplanets from residual diffracted starlight. 

The synergy of astrometric detection with \textit{Gaia} and \textit{WFIRST}-WFI and direct detection of exoplanets with SCALES has been identified with this simulation \cite{brandt2019astrometry}. Gaia’s extended 9-year survey is expected to yield a catalog of ~70,000 exoplanets \cite{Perryman2014gaia}. We predict 9 exoplanets ($<13 M_{jup}$) and 67 brown dwarfs ($>13 M_{jup}$) will be accessible to direct imaging with thermal emission by SCALES when initially detected through astrometry with \textit{Gaia} \cite{brandt2019astrometry}. With a single position measure by \textit{WFIRST}-WFI in 2030, this expands to 19 exoplanets and 144 brown dwarfs. These complementary measurements are sensitive to stellar mass estimates \cite{Nielsen2014} down to $\sim300$K, where water clouds manifest in M-band spectra \cite{skemer2016coldestbrowndwarfspectrum}. For previously-discovered directly-imaged planetary-mass or near-planetary-mass companions \cite{bowler2016}, SCALES will complement near-infrared spectroscopy with thermal infrared spectroscopy to improve estimation of important quantities for exoplanets, like luminosities, masses, molecular abundances (e.g. CO, CO${}_2$, H${}_2$O, CH${}_4$), temperatures, and cloud coverage.

For practical reasons, exoplanet imaging surveys have focused on young exoplanetary systems, where stellar emission is set by core mean molecular weight near pure hydrogen and exoplanets passively radiate residual heat of formation~\cite{Biller2007, vigan2012}. However, accretion shocks are efficient at radiating away energy in planets formed via core accretion, resulting in a population of exoplanets with relatively cold initial conditions ($<600$K). Near-infrared integral field spectrographs are not sensitive to such "cold-start" planets \cite{stone2018LEECH}, which have only imaged planets as cold as $\sim600-750$K \cite{Macintosh2015}. SCALES is sensitive to the thermal emission of this hypothetical population, probing the old and cold exoplanets.

In protoplanetary systems, it is possible to estimate mass accretion and $M_pM_d$ when young planets are embedded in natal discs \cite{Sallum2016, Currie2019}. Thermal infrared integral field spectroscopy would be particularly useful for distinguishing scattered light of circumstellar disks and regions of bright hydrogen-gas shocks consistent with protoplanets, as measured by the Br-$\gamma$ line. SCALES also uniquely opens the capacity to map circumstellar disks in water ice \cite{podio2013} and PAH emission. The simulation package is deployed with the previous examples included.

SCALES integral field spectroscopy also opens a unique parameter space in solar system, galactic and extragalactic astronomy. The contribution of background emission is diminished by dispersion, providing uniquely sensitive measurements of extended sources of thermal/redshifted emission. For example, volcanic eruptions and the extent of their lava fields can be investigated on Io (See Section 5.2), and carbon-based weather can be mapped on Titan. SCALES is sensitive to Brackett-, Pfund-, and CO-line mapping in bright young supernovae. The carbonaceous dust in unshocked ejecta of nearby remnants is sensitive to dust formation/destruction \cite{2005ApJ...628L.123K}. SCALES can also explore spatially-resolved nuclear/star-forming regions in nearby bright galaxies for characterization from PAH emission in dusty AGN tori \cite{2003A&A...398..101M} and hot dust emission.

\subsection{HR 8799-like System}

Lenslet-based integral field spectrographs are uniquely adept at imaging spectroscopy of exoplanetary systems because the optical distortion is not amplified when subsampling the focal plane with the lenslet array. The design of having the image slicer component downstream of the lenslet array also mitigates this problem that it would otherwise have, enabling high-spatial resolution thermal infrared integral field spectroscopy at low- and med- spectral resolutions.

Point sources were arrayed in the input field to be reminiscent of the HR 8799 system to represent an A0 star accompanied by three companions at 1000K, 1400K and 900K, each with radius 1 $R_J$, log(g) = 4.0, and distance = 40.9 pc. The PHOENIX atmospheric code was used to calculate the model spectra \cite{2011ApJ...733...65B}. The b component of the real system has no analog here, as is not contained within the field of view of the low-resolution integral field spectrograph when the other exoplanets are required to be in every image. 

\begin{figure}[h]
\includegraphics[width=13cm]{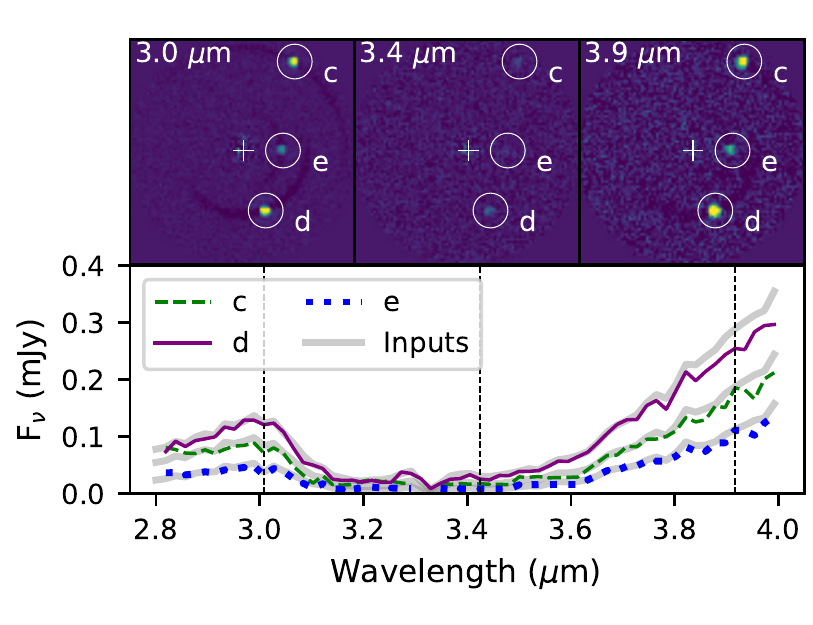}
\centering
\caption{Top: Simulations of the 3.0, 3.4 and 3.9 micron images of a 2-hr integration of a HR 8799-like system at $L$ band, processed with a naive, non-aggressive application of full-frame Angular Differential Imaging. Bottom: The input and recovered spectra of the HR 8799cde-like exoplanets from the same simulation at $L$ band}
\label{fig:fig6}
\end{figure}

Two hours of integration were simulated with positions and sky rotation set by starting when HR 8799 was nearest zenith. We used a constant phase screen and no coronagraph for internal testing purposes, and angular differential imaging was used to remove the point spread function of the primary. These observations are background limited for NIRC2 (with the possible exception of HR 8799 e) so this is a reasonable assumption until we implement the coronagraph in the code. Three representative slices of the data cube and the input/output spectra of this simulation are shown in Figure \ref{fig:fig6}.

\begin{figure}[h]
\includegraphics[width=16cm]{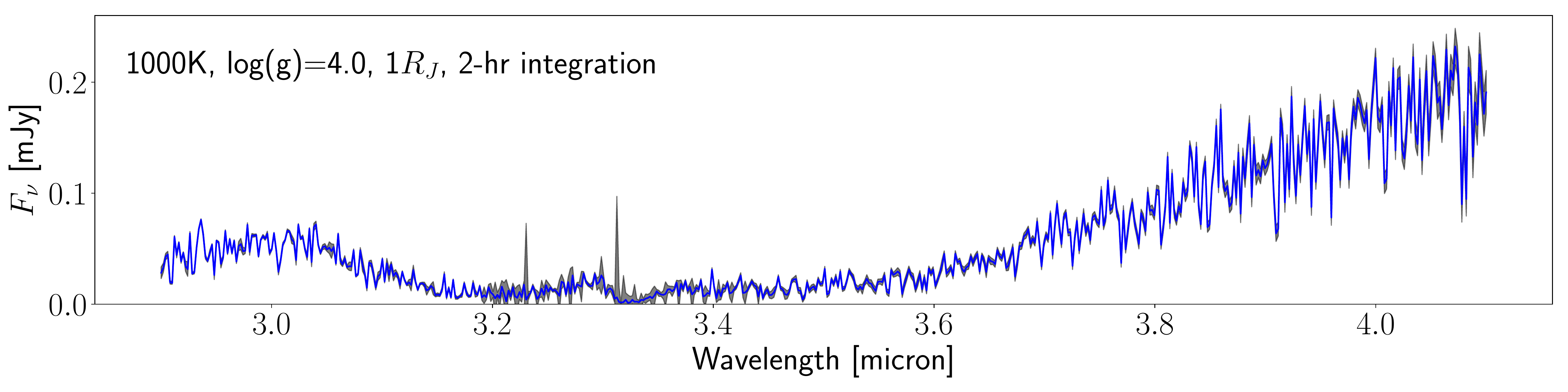}
\centering
\caption{The recovered spectrum of the analog of HR 8799c with error bars using the med-resolution mode in $L$ band for a 2-hr integration. The region near 3.3 microns has lower SNR due to both low telluric transmission and low astrophysical emission. Given the size of spectral resolution elements and linear dispersion model, there are 600 wavelength bins depicted here.}
\label{fig:fig7}
\end{figure}

For the med-resolution simulation, we look at a point source with an SED representative of HR 8799c. The small field of view of this mode requires an observation pattern that nods between the target, blank sky, and a PSF calibrator (usually the primary star). Two hours of total integration time following this pattern were simulated with positions and sky rotation set by starting when HR 8799 was nearest zenith (irrespective of whether that is consistent with daytime for the real HR 8799). The PSF was assumed not to translate due to pointing errors. The resulting recovered spectrum is depicted in Figure \ref{fig:fig7}.

\subsection{Io Volcanoes}

SCALES will monitor the locations, temperature, and extents of volcanoes on Io. For large eruptions, SCALES can map the extent of thermally active fields extending across the disc of Io while they cool. We assess SCALES's ability to recover the temperature of volcanoes on the surface of Io with a simple model for its surface. This model includes a uniform disk emitting isotropically with volcanoes as extended sources emitting at various temperatures enumerated in \cite{2017Natur.545..199D} to replicate lava fields as some volcanoes would be marginally resolved at the blue end of the 2-5 $\mu$m band. A more sophisticated model including Io's rotation as a uniform sphere emitting isotropically could be done in the future.

\begin{figure}[h]
\includegraphics[width=13cm]{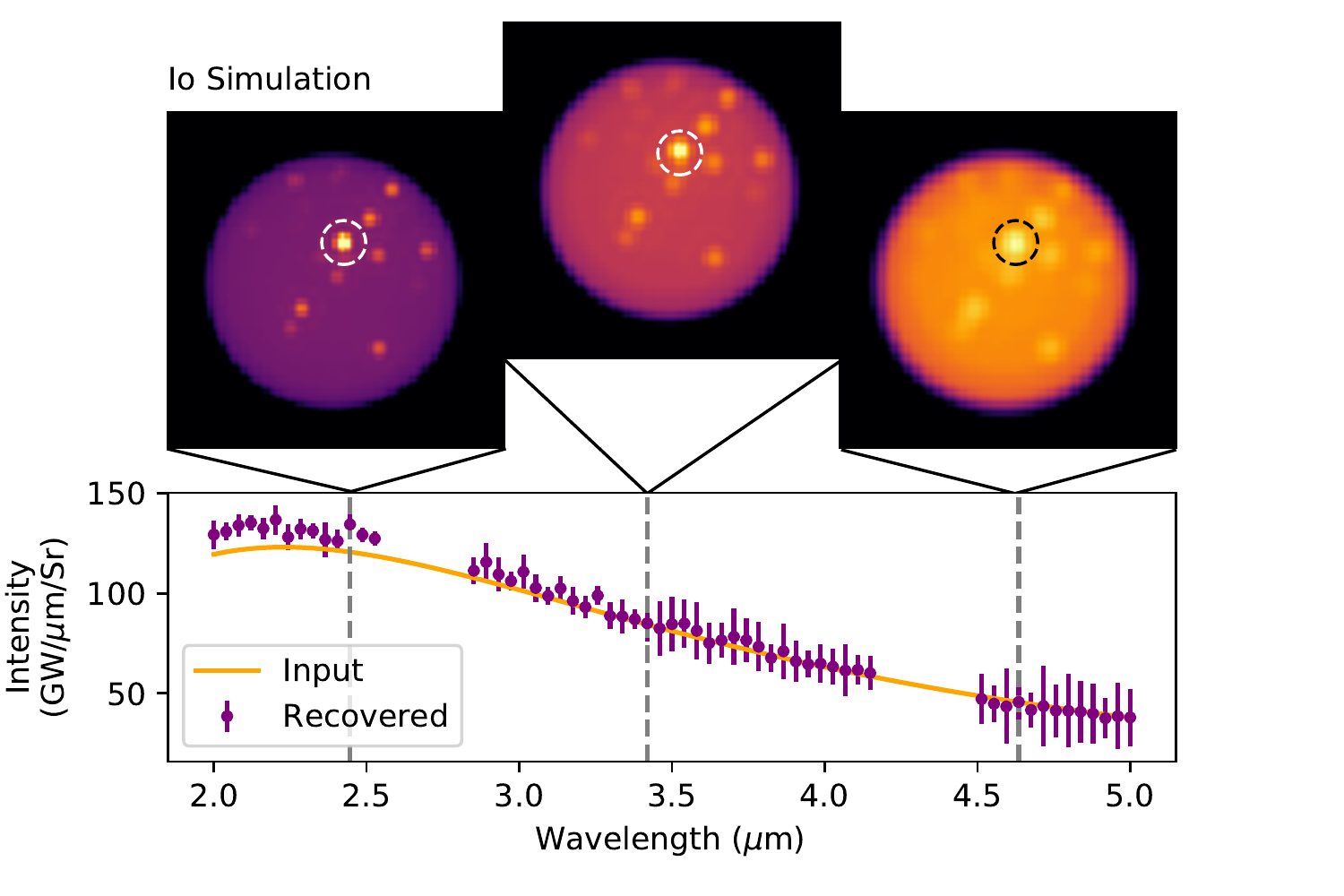}
\centering
\caption{Top: Simulations of the 2.4, 3.4 and 4.6 micron images of a 2-hr integration of our Io model at 2-5 $\mu$m band, processed with a naive, non-aggressive application of full-frame Angular Differential Imaging. Bottom: The input and recovered spectra of the Loki Patera-like volcano from the same simulation at 2-5 $\mu$m band. Note the excess intensity in $K$ band is caused by the volcano being marginally resolved and our extraction not accounting for this. The gaps in the spectrum are regions of high telluric absorption.}
\label{fig:fig8}
\end{figure}

We demonstrate the use of the 2-5 $\mu$m filter: simultaneous coverage of a significant portion of the spectral energy distribution from thermal emission of a volcano lava field. With one filter, we can determine the temperature of volcanoes (Figure \ref{fig:fig8}).

\section{Summary}

We presented the end-to-end simulation of the low-/med-resolution integral field spectrographs for SCALES, the dedicated high contrast IFU operating from 2-5 $\mu$m at R$\sim 100 - 10000$. Simultaneous development of the data reduction pipeline, simulation and the instrument itself enables exploration of consequences of design decisions and the capabilities of the instrument for delivering unique integral field spectrograph data.

\acknowledgments 
We gratefully acknowledge the support of the Heising-Simons Foundation through grant \#2019-1697. ZWB is supported by the National Science Foundation Graduate Research Fellowship under Grant No. 1842400. This paper is based on work funded by NSF Grant 1608834.

\bibliography{main} 
\bibliographystyle{spiebib}

\end{document}